\RequirePackage{ifpdf}
\ifpdf 
\documentclass[pdftex]{sigma}
\else
\documentclass{sigma}
\fi

\begin{document}

\allowdisplaybreaks

\renewcommand{\PaperNumber}{116}

\FirstPageHeading

\renewcommand{\thefootnote}{$\star$}

\ShortArticleName{Exact Solutions of the Equations of Relativistic
Hydrodynamics} 

\ArticleName{Exact Solutions of the Equations of Relativistic\\
Hydrodynamics Representing Potential Flows\footnote{This paper is
a contribution to the Proceedings of the Seventh International
Conference ``Symmetry in Nonlinear Mathematical Physics'' (June
24--30, 2007, Kyiv, Ukraine). The full collection is available at
\href{http://www.emis.de/journals/SIGMA/symmetry2007.html}{http://www.emis.de/journals/SIGMA/symmetry2007.html}}}

\Author{Maxim S. BORSHCH and Valery I. ZHDANOV}

\AuthorNameForHeading{M.S.~Borshch and V.I.~Zhdanov}

\Address{National Taras Shevchenko University of Kyiv, Ukraine}
\Email{\href{mailto:unacabeza@ukr.net}{unacabeza@ukr.net},
\href{mailto:zhdanov@observ.univ.kiev.ua}{zhdanov@observ.univ.kiev.ua}}
\URLaddress{\url{http://www.observ.univ.kiev.ua/astrophysics/zhdanov/}}

\ArticleDates{Received September 10, 2007, in f\/inal form
November 28, 2007; Published online December 07, 2007}

\Abstract{We use a connection between relativistic hydrodynamics
and scalar f\/ield theo\-ry to generate exact analytic solutions
describing
 non-stationary inhomogeneous f\/lows of the perfect f\/luid
with one-parametric equation of state (EOS) $p = p(\varepsilon)$.
For linear EOS $p = \kappa \varepsilon$ we obtain self-similar
solutions in the case of plane, cylindrical and spherical
symmetries. In the case of extremely stif\/f EOS ($\kappa=1$) we
obtain ``monopole~$+$~dipole'' and ``monopole~$+$~quadrupole''
axially symmetric solutions. We also found some nonlinear EOSs
that admit analytic solutions. }

\Keywords{relativistic hydrodynamics; exact solutions}

\Classification{76Y05; 83C15; 83A05}

\renewcommand{\thefootnote}{\arabic{footnote}}
\setcounter{footnote}{0}

\section{Introduction}\label{sec1}

Relativistic hydrodynamics (RHD) is extensively used to describe
various processes from mic\-roscopic to cosmological  scales.
Astrophysical applications of RHD deal with powerful
non-stationary phenomena such as the gamma-ray bursts and
hypernovae explosions. On the other hand, RHD solutions are
sources of hydrodynamical models of elementary particle multiple
production. Though at present there are powerful methods to f\/ind
numerical solutions of hydro\-dynamical problems, the price for
exact analytic representations of RHD f\/lows is still high.
However, the  nonlinear structure of the RHD equations hampers the
quest of analytic solutions and this is the reason why RHD
problems having exact solutions typically deal with simple
equation of state (EOS) such as $p = \kappa \varepsilon$ ($p$ is
the pressure, $\varepsilon$ is the proper frame energy density,
$\kappa ={\rm const}$). This linear equation of state\footnote{In
the case of the EOS $p = \kappa \varepsilon+B$ the constant $B$
can be easily eliminated from RHD equations by redef\/ining~$\varepsilon$.}
  is often used in multiple pion production theory since
works of Landau \cite{Landau} and Khalatnikov \cite{Khalatnikov}.

The most part of known exact RHD solutions concerns plane ($1+1$
dimensional) barotropic relativistic f\/lows. The Riemann simple
waves are the example (see, e.g.,~\cite{LandauLif}). In the case
of the linear EOS the method of Khalatnikov \cite{Khalatnikov}
allows to obtain a general  solution in an implicit form. However,
the solutions obtained by means of this method have rather
complicated representation. This motivated dif\/ferent authors to
look for more simple solutions to be used in models of multiple
pion production \cite{Hwa,Cooper,Chiu,Bjorken}; these solutions
have extensions to the case of spherical and cylindric symmetry.
Recent interest to such solutions is due to high energy heavy ion
collisions (see, e.g.,~\cite{Csorgo,Nagy,Pratt}).

The three-dimensional case of RHD is less developed. No general
analytic  solution is known in the case of relativistic
spherically symmetric f\/lows as well as in the case of cylindric
symmetry. The best known in theory of astrophysical relativistic
outbursts is the Blandford \& McKee approxi\-mate self-similar
solution \cite{BMC} that describes ultra-relativistic blast waves
of the f\/luid with the linear EOS; for modif\/ications and
generalizations (also in the ultra-relativistic approximation) see
\cite{Sari,ZhBor}.

Various techniques were invented in classical hydrodynamics in
order to f\/ind analytic solutions. In particular, one may
consider potential f\/lows of ideal f\/luid to reduce the number
of unknown functions. In this paper we study RHD f\/lows that may
be considered as a relativistic analog of the classical potential
f\/low: in these f\/lows the four-velocity is proportional to a
gradient of a scalar function. The relation between RHD f\/lows
and the scalar f\/ields is well known (see, e.g.,
\cite{Milekhin,Cooper,Korkina}) and at present it may be
considered as a part of physical folklore. This relation is
extensively used in cosmology in connection with the dark energy
problem \cite{Scherrer,Chimento,Sahni,Tsujikawa}. Here a number of
unusual equations of state has appeared; they have been considered
in the context of spatially homogeneous cosmological systems (see,
e.g., \cite{Tsujikawa} for the review).

In this paper we use the connection between RHD and the scalar
f\/ield theory that allows us to reduce the problem to an equation
for a single scalar function (instead of four-velocity and energy
density). The aim of this paper is to study the ability of this
trick to generate analytic RHD solutions that represent
non-stationary and non-homogeneous f\/lows of relativistic perfect
f\/luid. The paper is organized as follows. In Section~\ref{sec2} we
present basic equations and relations between the scalar f\/ield
and hydrodynamical variables. In Section~\ref{sec3} we present solutions
for an extremely stif\/f EOS: there are plane solutions, spherical
solutions including dif\/ferent combination of outgoing and
ingoing spherical waves; at last we present examples of
three-dimensional solutions that are obtained from spherical
harmonics of the scalar f\/ield. In Section~\ref{sec4} we present some
nonlinear equations of state that allow us to f\/ind exact RHD
solutions. In Section~\ref{sec5} we present two families of self-similar
solutions in the case of the linear EOS. In summary we brief\/ly
discuss the results obtained, their relation to the other known
solutions and possible generalizations.

\section{Notations and basic equations}\label{sec2}

The relativistic equations of the perfect f\/luid dynamics may be
written as conservation laws (e.g., \cite{Lichnerowicz}):
\begin{gather}
\label{eq1} \partial _{\nu } T^{\mu \nu } = 0,
\end{gather}
where
\begin{gather}
\label{eq2} T^{\mu \nu } = \left( {p + \varepsilon }
\right)\,u^\mu u^\nu - pg^{\mu \nu }, \qquad u^\mu u_\mu = 1,
\qquad g_{\mu \nu } = {\rm diag}\,\left( {1, - 1, - 1, - 1}
\right);
\end{gather}
$u^\mu $ is the four-velocity, $p$ is the pressure and
$\varepsilon $ is the proper frame energy density, the Greek
indices run from 0 to 3, the speed of light $c=1$. In the general
case equation~(\ref{eq1}) must be supplemented with the baryon
number conservation equations. However, it is not necessary in
this paper because we deal only with the one-parametric  EOS
$p=p(\varepsilon)$; provided EOS be known, the system~(\ref{eq1})
is complete. Further we concretize dif\/ferent equations of state
in every section.

If we have a conserved energy-momentum tensor of any f\/ield with
known solutions, we can use it to f\/ind a hydrodynamic solutions
for $\varepsilon$, $u^\mu $. However, to do this one needs to
represent this energy-momentum tensor in the form (\ref{eq2});
moreover, in order that these solutions represent physically
admissible motion of the relativistic f\/luid one must provide the
time-like behavior of the four-velocity. This is not always
possible. However, solutions with a spacelike $u^\mu $ are also of
some interest: they demonstrate that formal solutions of
completely relativistic dynamical equations may admit tachyonic
motions.

Transition to the proper frame of a local f\/luid element shows
that arbitrary tensor can be represented in the form (\ref{eq2}),
if the matrix $T^\mu _\nu $ has only one eigenvalue $\varepsilon >
0$ corresponding to a timelike eigenvector and triple-degenerate
eigenvalue $p$ with space-like eigenvectors. It is not always
possible to fulf\/ill these requirements for an arbitrary $T^\mu
_\nu $. However in the case of a scalar f\/ield with the
Lagrangian
\begin{gather}
\label{eq3} L = F(S),\qquad S = \frac{1}{2}\varphi _{,\alpha }
\varphi _,^{\;\alpha }
\end{gather}
\noindent the corresponding energy-momentum tensor
\begin{gather*}
 T_{\alpha \beta } = \varphi _{,\alpha } \varphi
_{,\beta } F'(S)- g_{\alpha \beta } F(S)
\end{gather*}
takes on the form (\ref{eq2}), if we put \cite{Milekhin}:
\begin{gather}
\label{eq4} p = F(S),\qquad \varepsilon = 2SF'(S) - F(S),
\\
\label{eq5} u_\alpha = \pm \varphi _{,\alpha } (2S)^{ - 1 / 2}.
\end{gather}
The equations (\ref{eq4}) provide an ef\/fective EOS that is
represented parametrically.

The equation of the scalar f\/ield following from (\ref{eq3}) is
\begin{gather}
\label{eq6}
\partial _\alpha [F'(S)\partial ^\alpha \varphi ] = 0 .
\end{gather}
Formulae (\ref{eq4}), (\ref{eq5}) represent an admissible
hydrodynamic f\/low provided that
\begin{gather}
\label{eq7} \varepsilon > 0,\qquad S > 0.
\end{gather}
This is the only condition to check, if we have a solution of
equation~(\ref{eq6}).

\section[Extremely stiff EOS: plane and spherical solutions]{Extremely stif\/f EOS: plane and spherical solutions}\label{sec3}

\subsection{Plane solutions}\label{sec3.1}

 The most simple is the case of a massless scalar f\/ield
 corresponding to  the extremely stif\/f EOS:
\begin{gather*}
F(S) = S,\qquad p = \varepsilon = S .
\end{gather*}
This EOS is considered throughout the whole section. In this case
we deal with a linear wave equation for the scalar f\/ield:
\begin{gather}
\label{eq9}
\partial _\mu \partial ^\mu \varphi = 0 .
\end{gather}
In the case of plane symmetry $(x^0 = t, \, x^1 = x)$ it is easy
to see that the general solution $\varphi = \psi \left( {t - x}
\right) + \chi \left( {t + x} \right)$ admits a hydrodynamics
interpretation if only
\begin{gather}
\label{cond1} \psi '\left( {t - x} \right)\chi '\left( {t + x}
\right)
> 0;
\end{gather}
this makes it evident that the solutions corresponding to the only
simple wave moving in one direction have no hydrodynamical
counterpart. For hydrodynamical  variables we have
\begin{gather}
\label{velocity}
 v = \frac{u^1}{u^0} = - \frac{\partial \varphi /
\partial x}{\partial \varphi / \partial t} = \frac{\psi \,'\left(
{t - x} \right) - \chi '\left( {t + x} \right)}{\psi \,'\left( {t
- x} \right) + \chi '\left( {t + x} \right)},\qquad \varepsilon =
2\psi \,'\left( {t - x} \right)\chi '\left( {t + x} \right).
\end{gather}
For $\psi '\left( x \right) = \chi '\left( x \right) = Ax^{ - 1}$
(throughout the paper $A={\rm const}\in {\mathbb R}$, $A\ne 0$) we
obtain the scaling solutions \cite{Hwa,Cooper,Chiu,Bjorken}: $v =
x / t$, $\varepsilon = 2A^2(t^2 - x^2)^{-1}$ that are well
def\/ined inside the light cone: $\vert x\vert < t$. For $x > t>0$
the hydrodynamical interpretation fails. If the solution for $t =
t_1 $, $\vert x\vert < x_1 < t_1 $ is considered as initial data
for further hydrodynamic evolution, either the solution must be
complemented by correct hydrodynamic data for $\vert x \vert > x_1
$, or the equations must be modif\/ied (cf., e.g.,
\cite{GorensteinZ,GorensteinS}) to extend the solution for all
$x$.
 Note also that after rescaling $A \to iA$ we obtain a
 formal ``tachyonic'' solution outside
the light cone with $\varepsilon > 0$, but with $v > 1$.

Choice $\psi '\left( x \right) = \chi '\left( x \right) =
Ax^\lambda$,
 $\vert x\vert < t$ (here and below $\lambda\in{\mathbb R}$, $\lambda\ne0$),
 generates similar solutions
with the energy density $\varepsilon = 2A^2(t^2 - x^2)^\lambda$
and $|v| < 1$. For $\lambda<0$ we have an outf\/low ($v>0$);  this
solution is singular at the light cone. For positive $\lambda$ we
have an inf\/low ($v < 0)$; in particular, for positive even
$\lambda$ the energy density has no light cone singularity. Simple
rescaling of factors in the functions $\psi$, $\chi $ generates
physical or tachyonic f\/lows correspondingly inside or outside
the light cone.

In order to construct a physical f\/low outside the light cone one
can choose $\chi '\left( x \right)=\psi '\left(- x
\right)=Ax^\lambda$, $ x > t>0$. We have an outf\/low for
$\lambda<0$, inf\/low for $\lambda>0$. This case can be easily
extended to negative $x$ and/or to negative $t$. For $\lambda=1$
we obtain $\varepsilon = 2A^2(x^2-t^2),\,v=-t/x$. This is a kind
of external scaling solutions discussed in the papers
\cite{Csorgo,Nagy}. In order to obtain strict inequality
(\ref{cond1}) and regular solutions $\forall\, x, \,t$, the above
power-law choice for $\psi$ and $\chi$ may be replaced, e.g., by
$\chi '\left( x \right)=\psi '\left(- x \right)\sim \exp(\gamma
x^n)$, where $\gamma={\rm const}\in{\mathbb R}$, $n$ is a positive
integer.


\subsection{Spherical solutions}\label{sec3.2}
The general solution in the case of spherically symmetry ($x^0 =
t,\;x^1 = r)$ is
\[
\varphi = r^{ - 1}\left[ {\psi \left( {t - r} \right) + \chi
\left( {t + r} \right)} \right],
\]
where $\psi $ and $\chi $ are arbitrary functions. Here, as
distinct from the planar case, it is possible to use the outgoing
wave solutions (with $\chi = 0)$:
\begin{gather}
\label{eq10} \varphi = \frac{\psi \left( {t - r} \right)}{r},
\end{gather}
whence
\begin{gather*}
p = \varepsilon = - \frac{\psi ^2}{2r^4}\left[ {1 + \frac{{2r\psi
}'}{\psi }} \right],
 \qquad v = 1 + \frac{\psi}{
r{\psi }'}.
\end{gather*}
In order to provide (\ref{eq7}), it is necessary
\begin{gather}
\label{eq11add}
 2\frac{{\psi }'\left( \alpha \right)}{\psi \left(
\alpha \right)}r < - 1,\qquad \alpha = t - r,
\end{gather}
\noindent therefore the function $\vert \psi \left( \alpha
\right)\vert $ is decreasing. Evidently it is impossible the
condition (\ref{eq11add}) to hold for all $t$, $r$ (in particular,
it is violated for $r \to 0$) and the hydrodynamical
interpretation of $T_{\mu \nu } $ corresponding to the solution
(\ref{eq10}) is possible only in a bounded region. When the sign
of $S$ changes and $\varepsilon = 0$ this solution must be matched
to vacuum in the way appropriate to hydrodynamic f\/low (see
Appendix A).

\medskip

\noindent \textbf{Example.} Consider a solution
\begin{gather*}
 \psi \left( \alpha \right) = C_1 \alpha ^{ - n} -
C_2,
\end{gather*}
$C_1>0$ and $C_2>0$ being real constants, $n$ is positive integer.
In this case (\ref{eq7}) is fulf\/illed if
\[
\psi + 2r{\psi }' = - \frac{\left[ {\left( {2n + 1} \right)r - t}
\right]C_1 }{\alpha ^{n + 1}} - C_2 < 0.
\]
This holds at least in the domain $\{(t,r):t > r > t / \left( {2n
+ 1} \right)\}$. As we see from Appendix A this solution is
matched to vacuum for $t - r = t_1 $, if $C_1 t_1^{ - n} = C_2 $.
For $r > t$ the hydrodynamical interpretation fails and it is
either necessary to match the solution through a discontinuity, or
the solution is destroyed by an external perturbation having
trajectory $r = - t + {\rm const}$.

\subsection{Combination of outgoing and ingoing spherical waves}
In this case, in order to provide regularity of the solution for
$r \to 0$, we put $\chi \left( t \right) = - \psi \left( t
\right)$
\begin{gather*}
\varphi \left( {t,r} \right) = r^{ - 1}\left\{ {\psi \left( {t -
r} \right) - \psi \left( {t + r} \right)} \right\}.
\end{gather*}
Consider the power-law choice of $\psi \left( x \right) = - Ax^{ -
n}$, $A > 0$, $n > 0$. For every $n$ the inequalities~(\ref{eq7})
must be analyzed separately. Below we present the hydrodynamic
solutions corresponding to four natural values of $n$; they have a
hydrodynamic interpretation inside the light cone $t > r$.

For $n = 1$, $t > r$:
\begin{gather}
\label{outin1} \varphi = - \frac{2A}{t^2 - r^2}, \qquad
\varepsilon = p = \frac{8A^2}{(t^2 - r^2)^3}, \qquad v =
\frac{r}{t}.
\end{gather}
Appropriate rescaling of the constant $A$ yields a tachyonic
solution outside the light cone.

For $n = 2$, $t > r$:
\begin{gather}
\label{outin2} \varphi = - \frac{4tA}{(t^2 - r^2)^2}, \qquad
\varepsilon = p = \frac{8A^2(9t^2 - r^2)}{(t^2 - r^2)^5}, \qquad v
= \frac{4tr}{3t^2 + r^2}.
\end{gather}

For $n = 3$: $t > r$:
\begin{gather}
\label{outin3} \varphi = - 2A\frac{3t^2 + r^2}{(t^2 - r^2)^3},
\quad \varepsilon = p = \frac{32A^2(9t^4 + 2t^2r^2 + r^4)}{(t^2 -
r^2)^7},\qquad v = \frac{r(5t^2 + r^2)}{3t(t^2 + r^2)}.
\end{gather}

For $n = 4$, $t > r$:
\[
 \varphi = - 8A\frac{t(t^2 + r^2)}{(t^2 - r^2)^4},
\]
\begin{gather}
\label{outin4}
 \varepsilon = p =
\frac{32A^2[10t^4(t^2+r^2)+15t^2(t^4+t^2r^2+r^4)- r^6]}{(t^2 -
r^2)^9}, \qquad v = \frac{2tr(3r^2 + 5t^2)}{5t^4 + 10t^2r^2 +
r^4}.
\end{gather}

In a more general case $( n \ge 1$, $A > 0 )$ one can show that
hydrodynamic f\/low is correctly def\/ined at least in the domain
$\{(t,r): \kappa _n t < r < t\}$, where $\kappa _n = (a_n - 1)/(
a_n + 1 )$, $a_n = \left[ 2(n + 1) \right]^{1/n} $, and in this
region $u^1 > 0$.

\section[Extremely stiff EOS: solutions based on spherical harmonics]{Extremely stif\/f EOS: solutions based on spherical harmonics}\label{sec4}

In this section we proceed with the extremely stif\/f EOS; here
generation of RHD solutions utilizes the well-known formula for
the solution of the wave equation (\ref{eq9}) using spherical
functions $Y_{lm} \left( {\theta ,\varphi } \right)$ (in spherical
coordinates)
\begin{gather}
\label{eq19} \varphi \left( {t,r,\theta ,\varphi } \right) =
\sum\limits_{l = 0}^\infty {\;\sum\limits_{m = - l}^l {\;\tilde
{f}_{lm} \left( {t,r} \right)Y_{lm} \left( {\theta ,\varphi }
\right)} },
\\
\label{eq19a}
 \tilde {f}_{lm} = r^l\left(
{\frac{1}{r}\frac{\partial }{\partial r}} \right)^l\left\{
{\frac{f_{lm}^{\left( 1 \right)} \left( {t - r} \right) +
f_{lm}^{\left( 2 \right)} \left( {t + r} \right)}{r}} \right\},
\end{gather}
where $f_{lm}^{\left( 1 \right)} \left( {t - r} \right)$,
$f_{lm}^{\left( 2 \right)} \left( {t + r} \right)$ are arbitrary
functions.

Investigation of inequality (\ref{eq7}) for (\ref{eq19}) is rather
complicated in the general case. However, new special solutions
obeying (\ref{eq7}) may be easily obtained as follows. Take one of
spherically
 symmetrical solutions (\ref{outin1})--(\ref{outin4})
from the previous section and add some items from the sum in the
r.h.s.\ of~(\ref{eq19}). These items must be suf\/f\/iciently
small in comparison with (\ref{outin1})--(\ref{outin4}) to
preserve inequality (\ref{eq7}) at least within a compact domain
inside the light cone. This is possible for a special choice of
appropriate functions (\ref{eq19a}). Below we present two such
special solutions that represent a hydrodynamical f\/low within
the light cone $t>r$.

In the case of axial symmetry we obtain the three-dimensional
velocity components from the relations
\[
v^r \equiv \frac{u^r}{u^0} = - \frac{\varphi _{,r} }{\varphi _{,t}
},\qquad v^\theta \equiv \frac{u^\theta }{u^0} =
-\frac{1}{[r\sin(\theta)]^2}\frac{\varphi _{,\theta } }{\varphi
_{,t} }.
\]

The next case corresponds to the monopole + dipole contribution
into $\varphi $:
\[
\varphi = A\frac{t - br\cos \theta }{(t^2 - r^2)^2},
\]
$b$ is a real constant. If $| b | < 1$, then the condition
(\ref{eq7}) is satisf\/ied for all $t > r$. Corresponding solution
in terms of hydrodynamic variables is
\begin{gather}
\label{mondip1}
 \varepsilon = p = \frac{A^2}{2(t^2 - r^2)^5}\left[
{(1 - b^2)(t^2 - r^2) + 8\left( {t - br\cos \theta } \right)^2}
\right]
> 0,
\\
\label{mondip2}
v^r = \frac{4tr - (3r^2 + t^2)b\cos (\theta )}{3t^2 + r^2 -
4trb\cos (\theta )}, \qquad v^\theta = \frac{(t^2 - r^2)b\sin
(\theta )}{r[3t^2 + r^2 - 4trb\cos (\theta )]}.
\end{gather}

The other solution is generated by a monopole $+$ quadrupole
contribution
\[
\varphi = A\frac{3t^2 + r^2 + br^2P_2 }{(t^2 - r^2)^3},
\]
 where $P_2 = [3\cos ^2(\theta ) - 1] / 2$, $b$ is a real
constant. We checked that the condition (\ref{eq7}) is also
satisf\/ied in this case for all $t > r$, at least if $|b| < 1$.
Corresponding hydrodynamical solution is
\begin{gather}
 \varepsilon = p = \frac{A^2}{2(t^2 - r^2)^{
7}}\big[ {144t^4 + 16r^2(2t^2 + r^2)\left( {1 + 2bP_2 } \right) }
\nonumber\\
\phantom{\varepsilon = p =}{} -  {2r^2b^2(P_2 + 1)(t^2 - r^2) +
12b^2P_2^2 r^4} \big] >
0,\label{quadip1}
\\
\label{quadip2}
 v^r = \frac{r}{3t}   \frac{10t^2 + 2r^2 + (t^2
+ 2r^2)bP_2 }{2t^2 + 2r^2 + br^2P_2 }, \qquad v^\theta = -
\frac{b}{2t}   \frac{(t^2 - r^2)\sin (\theta )\cos (\theta )}{2t^2
+ 2r^2 + br^2P_2 }.
\end{gather}

\section{Nonlinear barotropic EOS}\label{sec5}

In this section we consider equation~(\ref{eq6}) in the case of
plane ($n=0$), cylindrical ($n=1$) and spherical ($n=2$) symmetry.
The dependence $p=p(\varepsilon)$ will be specif\/ied later. In
this case equation~(\ref{eq6}) can be written as
\begin{gather}
\label{eq20a} \frac{\partial }{\partial t}\left(
{F'(S)\frac{\partial \varphi }{\partial t}} \right) =
\frac{1}{r^n}\frac{\partial }{\partial r}\left( {r^n F'(S)
\frac{\partial \varphi }{\partial r}} \right).
\end{gather}
\noindent  We are looking for solutions of the equation
(\ref{eq6}) of the form
\begin{gather}
\label{eq21} \varphi=\varphi(\sigma), \qquad \sigma=(t^2-r^2)/2
\end{gather}
In this case
\[
S=\sigma \left( \frac{d \varphi}{d \sigma}\right)^2
\]
and substitution into equation~(\ref{eq20a}) yields
\begin{gather}
\label{eq22} 2\sigma \frac{d }{d \sigma}\left[ F'(S)\frac{d
\varphi}{d \sigma} \right]+(2+n)F'(S)\frac{d \varphi}{d \sigma}=0.
\end{gather}
It is convenient to introduce a new variable $\tau$ by the
relation $\sigma=\tau^2/2$; $S=(1/2)(d\varphi/d\tau)^2$. Then
equation~(\ref{eq22}) yields
\begin{gather}
\label{eq23}  {\tau^{1+n}F'(S)}\frac{d \varphi}{d \tau} ={\rm
const}.
\end{gather}

We investigate the most simple cases, when equation~(\ref{eq23})
can be easily solved with respect to $d\varphi / d\tau  $ and the
 EOS is expressible in terms of elementary functions.

Consider f\/irst the EOS
\begin{gather}
\label{eq24}
p=\varepsilon\left[\ln\left(\frac{\varepsilon}{\varepsilon_0}\right)+B\right],\qquad
F(x^2/2)=\varepsilon_0 x[\ln(x)+B],
\end{gather}
\noindent where $B$ is a dimensionless constant, $ \varepsilon_0$
is a constant having the dimension of the energy density.
 Substitution into
equation~(\ref{eq23}) gives us
\[
\frac{d \varphi}{d \tau}=\exp\left(\frac{C}{ \tau^{1+n}}-B-1
\right),
\]
$C$ being an arbitrary real constant. The solution of
hydrodynamical equations with EOS (\ref{eq24}) is
\begin{gather}
\label{eq24a}
 v=r/t, \qquad  \varepsilon=\varepsilon_0 \exp\left(\frac{C}{
\tau^{1+n}}-B-1 \right), \qquad  \tau=\sqrt{t^2-r^2}, \qquad t>r.
\end{gather}
This solution is formally valid for interior of the future light
cone. However, for usual hydrodynamical interpretation we
need\footnote{See, however, remarks in the last section.} that
\begin{gather}
\label{sound}
 c_s^2=dp/d\varepsilon\in [0,1],
\end{gather}
where $c_s$ is the speed of sound. This yields an additional
restriction of the domain of RHD solution: $e^{-1}\le \varepsilon
e^B/\varepsilon_0\le 1$ whence $t^2-r^2\ge C^{2/(1+n)}$, $C\ge0$.

 In a more complicated case we consider
\begin{gather*}
 F(S)=\varepsilon_0 \sqrt{S}\sum\limits_{m = 0}^N
\frac {(-1)^m N!}{(N-m)!}\left(\ln \sqrt{S}\right)^{N-m}
\end{gather*}
with
\[
\frac{dF}{dS}=\frac {\varepsilon_0}{2\sqrt{S}} \left(\ln
\sqrt{S}\right)^{N}.
\]
Corresponding EOS is represented parametrically ($R=\sqrt S$) as
\begin{gather}
\label{eq24b}
 \varepsilon=- \varepsilon_0 R\sum\limits_{m = 1}^N \frac
{(-1)^m N!}{(N-m)!}\left(\ln R\right)^{N-m}, \qquad
p=\varepsilon_0 R\sum\limits_{m = 0}^N \frac {(-1)^m
N!}{(N-m)!}\left(\ln R\right)^{N-m},
\end{gather}
and both $\varepsilon$ and $p$ may be expressed as functions of
enthalpy
\[
\varepsilon+p=\varepsilon_0 R\left(\ln R\right)^{N}.
\]
 In this case
\begin{gather}
\label{csN} c_s^2=\frac{dp}{d\varepsilon}=\frac{\ln R}{N}
\end{gather}
On account of equation (\ref{eq23})
\[
\tau^{1+n}\left(\ln R\right)^N ={\rm const}.
\]
This  yields the hydrodynamical f\/low for the EOS (\ref{eq24b})
with
\begin{gather}
\label{nonlin2}
 v=r/t, \qquad  R=\exp\left[C
(t^2-r^2)^{-(1+n)/(2N)}\right],  \qquad 0<r<t,
\end{gather}
$C $ is a real constant. Taking into account of
equations~(\ref{csN}),
 (\ref{sound}) yields
\begin{gather*}
 1\le R\le e^N,\qquad t^2-r^2\ge
(C/N)^{2N/(1+n)},\qquad C\ge 0.
\end{gather*}

\section[Linear EOS $p = \kappa \varepsilon$]{Linear EOS $\boldsymbol{p = \kappa \varepsilon}$}\label{sec6}

It is easy to f\/ind the Lagrangian (\ref{eq3}) corresponding to
the EOS $p = \kappa \varepsilon $, where $\kappa = c_s^2$, $0 <
\kappa < 1$. Using (\ref{eq4}) we obtain a simple dif\/ferential
equation for $F$ yielding
\begin{gather}
\label{eq27add}
 F(S) = S^\alpha ,\qquad \alpha = \frac{1 + \kappa}{2\kappa },
\end{gather}
 up to inessential multiplier.

Such Lagrangians and their generalizations have been discussed in
\cite{Cooper,Scherrer,Chimento,Grigoriev}. Now we consider
equation~(\ref{eq6}) in the case of plane, cylindrical and
spherical symmetry on account of equation~(\ref{eq27add}).
Equation~(\ref{eq20a}) is reduced to the following equation
\begin{gather}
\label{eq25} \frac{\partial }{\partial t}\left( {S^{\alpha -
1}\frac{\partial \varphi }{\partial t}} \right) =
\frac{1}{r^n}\frac{\partial }{\partial r}\left( {r^nS^{\alpha -
1}\frac{\partial \varphi }{\partial r}} \right).
\end{gather}

First, we are looking for the solutions of the form (\ref{eq21}).
We have a special case of equation~(\ref{eq23}) that can be easily
solved yielding the famous scaling solutions of relativistic
hydrodynamics \cite{Hwa,Cooper,Chiu,Bjorken}:
\begin{gather*}
\label{scaling} \varepsilon=\frac{C}{\tau^{(1+k)(1+n)}}, \qquad
v=r/t, \qquad 0<r<t, \qquad C>0.
\end{gather*}

Now we shall look for solutions of equation~(\ref{eq25}) of the
form $\varphi = \varphi (\xi ),\quad \xi = r / t$. Then
\[
S=\frac{( \xi^2-1)}{2 t^2} \left( \frac{d \varphi }{d \xi
}\right)^2 .
\]
One can check by direct calculations that equation~(\ref{eq25})
leads to the equation
\begin{gather*}
\label{dzdxi}
 \frac{dz}{d\xi } = \frac{nz}{\xi (\xi ^2 -
1)},\qquad z = \left[ {(\xi ^2 - 1)\frac{d\varphi }{d\xi }}
\right]^{2\alpha - 1}.
\end{gather*}

For $\xi > 1$ we obtain
\begin{gather*}
\frac{d\varphi }{d\xi } = \frac{C}{\xi ^{nk}(\xi ^2 - 1)^{1 - nk /
2}}, \qquad C={\rm const}.
\end{gather*}
The hydrodynamic variables are as follows
\begin{gather}
\label{similar} v = \frac{t}{r},\qquad p = \kappa \varepsilon  =
S^\alpha , \qquad S = \frac{C^2}{2r^{2n\kappa }(r^2 - t^2)^{1 -
n\kappa }};
\end{gather}
the solution is valid outside the light cone ($0<t < r$). Again,
after rescaling of the constant $C$ we have a tachyonic solution
inside the light cone with $\varepsilon
> 0$, $\vert v\vert > 1$.

For $n=1,2$ there is the  light-cone singularity for all
$k\in(0,1)$. The same situation occurs in the case of $k<1/2$ for
$n=2$. In the case of a continuous f\/low the solution
(\ref{similar}) may be matched with a numerical solution through a
sound wave.

\begin{figure}[t]
\centerline{\includegraphics[width=2.4 in]{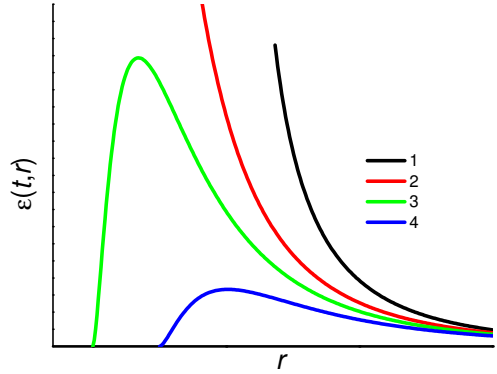}}
 \caption{Energy density $\varepsilon(t,r)$ according
to equation~(\ref{similar}), $n=2$, for dif\/ferent times $t=t_i$
in the case of $kn<1$ (lines 1, 2: $t_1>t_2$) and $kn>1$ (lines 3,
4: $t_3<t_4$).\label{fig1} }
\end{figure}

In the case of spherical symmetry ($n=2$) and $k>1/2$ the energy
density has no light cone singularity and we have a continuous
solution (\ref{similar}) that is matched with vacuum at the light
cone (Fig.~\ref{fig1}); this may be checked using
equation~(\ref{eq13}) from the Appendix~\ref{appendixA}. In the case of $kn=1$
the density does not depend on time describing an outf\/low
supported by an energy source from the interior.

\section{Summary and discussion}\label{sec7}
In this paper we considered relativistic potential f\/lows of the
perfect f\/luid that permit reductions to an equation for one
scalar f\/ield. The restriction to the potential f\/low allowed us
to derive analytic RHD solutions that are interesting from
theoretical viewpoint, e.g., in the theory of elementary particle
multiple production; they may be used  to test numerical RHD
computer routines. The relation between RHD and the scalar f\/ield
proved to be most fruitful in the case of the extremely stif\/f
EOS. In this case the problem is reduced to the linear wave
equation for the scalar f\/ield yielding hydrodynamical solutions
in the cases of planar, cylindrical and spherical symmetry.
Moreover, a family of ``non-symmetric'' RHD solutions may be
generated using multipole expansion for the scalar f\/ield; this
is illustrated by examples that arise from combination of lower
multipoles. We also found a family of nonlinear EOSs that permit
analytic solutions def\/ined inside the light cone and having
behavior like that of famous scaling solutions
\cite{Hwa,Cooper,Chiu,Bjorken}. In the case of the linear EOS $p =
\kappa \varepsilon$ some of our solutions reduce to these
solutions \cite{Hwa,Cooper,Chiu,Bjorken}. Also, we derived the
solutions outside the light cone from self-similar solutions of
the nonlinear scalar f\/ield equation.

Most of our solutions have light cone singularities like that of
\cite{Hwa,Cooper,Chiu,Bjorken}. These solutions cannot be directly
matched with vacuum and may be used to describe a f\/low in a
f\/inite volume until it is destroyed by perturbations moving from
the boundaries. Nevertheless, there are exceptions when regular
solutions in a whole space may be generated, especially in the
case of $k=1$ and in the case of spherical solutions
(\ref{similar}) with $k>1/2$.

After we have sent the f\/irst version of our manuscript to the
journal the paper \cite{Nagy} was published\footnote{We are
thankful to the anonymous referee for drawing our attention to the
papers \cite{Csorgo,Nagy}.} containing new interesting RHD
solutions. Some of the solutions obtained in the present paper
were independently found in \cite{Nagy} using dif\/ferent methods.
 In particular, our solution (\ref{similar}) corresponds to
equations~(24), (25) of~\cite{Nagy} that have been obtained as
 an extension of ``accelerating'' solutions of~\cite{Csorgo}. Also, our solution~(\ref{velocity}) in the case of the plane f\/low and extremely stif\/f EOS is the
 same as solution (114), (115) of~\cite{Nagy}.

To outline possible generalizations, we note that the condition
about the speed of sound ($0<k<1$ in the case of the linear EOS or
equation~(\ref{sound}) of Section~\ref{sec5}) may be relaxed. This may be
of some physical interest: the EOS with negative pressure,
negative $k<0$ or $k>1$ are discussed in cosmology
\cite{Sahni,Tsujikawa}, see also \cite{Babichev} concerning the
possibility of ``superluminal'' EOS. The results of the Section~\ref{sec5}
will be extended, if we do not restrict ourselves to simplest EOS
in terms of elementary functions. In Section~\ref{sec6} we conf\/ine
ourselves to the self-similar solutions of the f\/ield
equation~(\ref{eq25}) of the form $\varphi=\varphi(r/t)$ thus
yielding an ordinary dif\/ferential equation. However, more
general substitution $\varphi=r^{\alpha}\psi(r/t)$ also leads to
an ordinary dif\/ferential equation, though more complicated.

\appendix

\section{Matching of solutions with vacuum}\label{appendixA}
To perform  this matching  in a hydrodynamical way we use the
condition of zero energy-momentum f\/lux through a boundary with
vacuum
\begin{gather}
\label{eq13} T_\mu ^\nu k_\nu = 0,
\end{gather}
\noindent where the vector $k_\nu $ is orthogonal to the boundary
hypersurface. This yields
\begin{gather*}
T_{\mu \nu } k^\nu = \varphi _{,\mu } \varphi _{,\nu } k^\nu -
\frac{1}{2}k_\mu \left( {\varphi _{,\alpha } \varphi _, ^\alpha }
\right) = 0.
\end{gather*}
Simple analysis shows that this is possible only if at the
boundary
\begin{gather*}
\varphi _{,\alpha } \varphi _, ^\alpha = 0 , \qquad \varphi
_{,\alpha } k^\alpha = 0 .
\end{gather*}
In the two-dimensional case we have $\varphi _{,0}^2 = \varphi
_{,1}^2 $.

Consider the light-like surface $t - r - t_1 = 0 $, $\alpha = t_1
$, then $k^0 = k^1 = - k_1 = 1$. In the case of the solution
(\ref{eq10}) the boundary condition is
\begin{gather*}
T_{\mu \nu } k^\nu = \frac{\psi ^2\left( \alpha
\right)}{2r^4}k^\mu ,
\end{gather*}
\noindent that is at the boundary with vacuum $\psi \left( {t_1 }
\right) = 0$.

\subsection*{Acknowledgements}
 We are thankful to the referees of our paper
for helpful remarks and suggestions. This work has been supported
in part  by ``Cosmomicrophysica'' program of National Academy of
Sciences of Ukraine.

\pdfbookmark[1]{References}{ref}
\LastPageEnding

\end{document}